\documentclass[twocolumn,usenatbib,amsmath,amssymb,amsbsy]{mn2e}

\usepackage{natbib}
\usepackage{amsbsy}
\usepackage{amssymb}
\usepackage{amsmath}

\usepackage{graphicx}

\newcommand{\be}{\begin{equation}}
\newcommand{\ee}{\end{equation}}
\newcommand{\bear}{\begin{eqnarray}}
\newcommand{\eear}{\end{eqnarray}}
\newcommand{\rx}{{\rm x}}
\newcommand{\re}{{\rm e}}

\newcommand{\rc}{{\rm c}}

\newcommand{\rn}{{\rm n}}
\newcommand{\rp}{{\rm p}}
\newcommand{\rnp}{{\rm np}}
\newcommand{\rpn}{{\rm pn}}

\newcommand{\rpe}{{\rm pe}}

\newcommand{\rs}{{\rm s}}
\newcommand{\rns}{{\rm ns}}

\newcommand{\rP}{{\rm P}}
\newcommand{\rsf}{{\rm sf}}

\newcommand{\mut}{{\tilde{\mu}}}

\newcommand{\Om}{\Omega}
\newcommand{\dmut}{{\delta \mut}}
\newcommand{\xp}{{x_\rp}}
\newcommand{\cR}{{\cal R}}

\newcommand{\cB}{{{\cal B}}}
\newcommand{\cBp}{{{\cal B}^\prime}}

\newcommand{\cD}{{\cal D}}

\newcommand{\cF}{{\cal F}}
\newcommand{\wns}{\vec{w}^{\,\rm{ns}}}
\newcommand{\ws}{\vec{w}^{\, \rm{s}}}
\newcommand{\wss}{w^{\rm{s}}}
\newcommand{\wnss}{w^{\rm{ns}}}
\newcommand{\taus}{\tau^{\rm{s}}}
\newcommand{\tauns}{\tau^{\rm{ns}}}
\newcommand{\tausP}{\tau^{\rm{s}}_\rP}
\newcommand{\tausphi}{\tau^{\rm{s}}_\varphi}
\newcommand{\taunsP}{\tau^{\rm{ns}}_\rP}
\newcommand{\taunsphi}{\tau^{\rm{ns}}_\varphi}

\begin{document}

\title[Ambipolar diffusion in superfluid neutron stars] {Ambipolar diffusion in superfluid neutron stars}

\author[Glampedakis,  Jones \& Samuelsson]{K. Glampedakis$^1$,  D.I. Jones$^2$ \&  L. Samuelsson$^{3, 4}$\\
  \\
  $^1$  Theoretical Astrophysics, University of T\"ubingen, Auf der Morgenstelle 10, T\"ubingen, D-72076, Germany \\
  $^2$ School of Mathematics, University of Southampton, Southampton
  SO17 1BJ, UK \\
  $^3$ Department of Physics, Ume\aa\ University, SE-901 87 Ume\aa, Sweden \\
$^4$ Nordita, Roslagstullsbacken 23, SE-106 91 Stockholm, Sweden }

\maketitle

\begin{abstract}

In this paper we reconsider the problem of magnetic field diffusion in neutron star cores.  We model the star as consisting of a 
mixture of neutrons, protons and electrons, and allow for particle reactions and binary collisions between species.  Our analysis 
is in much the same spirit as that of Goldreich \& Reisenegger (1992), and we content ourselves with rough estimates of magnetic 
diffusion timescales, rather than solving accurately for some particular field geometry.  However, our work improves upon previous 
treatments in one crucial respect: we allow for superfluidity in the neutron star matter.  We find that the consequent mutual friction 
force, coupling the neutrons and charged particles, together with the suppression of particles collisions and reactions, drastically 
affect the ambipolar magnetic field diffusion timescale.  In particular, the addition  of superfluidity means that it is unlikely that 
there is ambipolar diffusion in magnetar cores on the timescale of the lifetimes of these objects, contradicting an assumption often made 
in the modelling of the flaring activity commonly observed in magnetars.  Our work suggests that if a decaying magnetic field is 
indeed the cause of magnetar activity, the field evolution is likely to take place outside of the core, and might represent Hall/Ohmic 
diffusion in the stellar crust, or else that a mechanism other than standard ambipolar diffusion is active, e.g. flux expulsion due to 
the interaction between neutron vortices and magnetic fluxtubes. 
 
\end{abstract}

\begin{keywords}
stars: magnetars -- stars: neutron -- stars: magnetic fields
\end{keywords}

%%%%%%%%%%%%%%%%%%%%%%%%%%%%%%%%%%%%%%%%%%%%%%%%%%%%%%%%%%%%%%%%%%%%%%%%%%%%%%%%%%%

\section{Introduction}

\label{sec:intro}

The structure and evolution of magnetic fields in compact stars is of considerable interest.  In part this is due to the large amount of 
information available: magnetic field strengths have been measured, or more often inferred, for many neutron stars, allowing astronomers 
to build up a detailed picture of the distribution of fields strengths in them.  Crucially, there seem to be pronounced 
systematic differences in field strength between different classes of neutron stars.  The young neutron stars and most other `normal' 
pulsars have fields of the order $10^{12}$ G, while the older millisecond pulsars (MSPs) have much lower fields, of order $10^9$ G.  
Similarly (relatively) low fields are believed to exist in the neutron stars located in accreting low-mass X-ray binary systems (LMXBs).  
In contrast, magnetars, as observed as anomalous X-ray pulsars (AXPs) and soft gamma repeaters (SGRs), seem to have exceptionally 
high field strengths, of order $10^{15}$ G.  

The way in which magnetic fields slowly evolve in neutron stars may well be crucial in understanding the relationship between 
these classes of objects. There are indications that millisecond pulsars acquire their relatively low field strength through 
a period of accelerated field decay while accreting in LMXBs, neatly linking the normal, LMXB and MSP populations.

Even more intriguingly, the structure, strength and, most of all, the evolution of the magnetic field may provide a unique insight 
into the state of the stellar interior. Field decay in the hot crust of an LMXB has been invoked to explain the reduction in field 
strength that accretion is believed to bring about, giving information on the conductivity of the crustal matter (see e.g. \citet{caz04}). 
More exotically, magnetic field evolution has been invoked to explain the restless behaviour of the magnetars. As was argued convincingly by 
\citet{TD95,TD96,TD01}, the existence of a $10^{15}$ G strength field explains many of the observed features of the AXPs and the SGRs.  
In particular, Thompson and Duncan pointed out that the very slow rotation rates of the magnetars imply that the energy required 
to power their emission cannot be supplied by rotation, but a $10^{15}$ G magnetic field, evolving on the inferred lifetime of the 
magnetars (of order $10^3-10^4$ years), would provide a suitable energy reservoir. 

In order for this model to be self-consistent, there must exist a mechanism capable of causing a $10^{15}$ G magnetic field to evolve 
on a timescale of order $10^4$ years.  The issue of magnetic field evolution had been considered previously by \citet{GR92} 
(hereafter GR92), who showed that a variety of processes contribute to the field evolution in a neutron star. 
These processes are (i) Ohmic decay, in which the current and therefore the field decays through the frequent collisions between the 
different particle species; 
(ii) Hall drift, in which the magnetic field produced by the current acts back on the current itself, producing a (dissipation free) 
rearrangement of the magnetic flux; (iii)  ambipolar drift, in which the charged particles (electrons and protons) and the magnetic 
field move relative to the neutral fluid (the neutrons).  It was the last of these, ambipolar drift, that Thompson and Duncan were 
able to invoke to explain the activity of the magnetars.   While not important for normal field strength stars, they noted that 
the quadratic scaling of the drift velocity with magnetic field strength led to field rearrangement timescales of order magnetar 
lifetimes for magnetar-like field strengths.
In contrast, Ohmic decay is unlikely to be important in neutron star cores \citep{baym}.
The Hall drift, although not a dissipative process itself, might cause a rearrangement of the 
magnetic field on astrophysically relevant timescales, assuming a magnetar-strong magnetic field (GR92).
However, as we discuss in this paper, this effect is likely to be suppressed in the presence
of proton superconductivity. It should be noted, nevertheless, that the possibility remains that Hall and Ohmic processes, 
working in concert, might lead to significant field evolution in the stellar crust; see e.g. \citet{pmg09}.

It is precisely this issue of magnetic field evolution in a neutron star core that we revisit in this paper.  
Given its apparent significance for the magnetar model, we pay particular attention to ambipolar diffusion. Our starting point is the 
classic work of Goldreich and Reisenegger (GR92), who modelled 
neutron stars as being composed of a neutron-proton-electron plasma, with field evolution being driven by binary scattering processes 
between the species.  The analysis of GR92 was essentially at the level of estimating the key timescales, rather than calculating explicit 
fully consistent solutions.  Their analysis has recently been extended by \citet{hoyos08,hoyos10}, in which the magnetic evolution problems 
were solved explicitly in a simplified geometry, to monitor the validity of the timescales estimated by GR92, and to understand the change 
in geometry brought about by the diffusive processes.

Our aim in this paper is somewhat different.  Rather than obtaining more explicit solutions to the original model, we instead will content 
ourselves with making simple estimates in the spirit of GR92, but with the addition of the crucial ingredient of neutron superfluidity.  
There is strong evidence, both from theory and observation, that mature neutron stars are likely to be sufficiently cold that the nuclear 
matter is in a superfluid state, see e.g. \citet{sauls}. Our aim is to see what qualitative features result from this more realistic level of 
modelling, and, in particular, to see if timescales of order of the magnetar ages naturally emerge. That the timescales might change 
significantly is plausible, given that superfluidity is known to suppress particle collisions and reactions, but introduce a new mechanism for 
coupling the different fluids, namely mutual friction, in which the neutron and charged fluids are effectively coupled.  This coupling might be 
brought about by the scattering of electrons off the vortices present in the rotating superfluid \citep{als88}.  It is the changes in the field 
evolution timescales, and their implications for the magnetar model, that we seek to investigate.

The structure of this paper is as follows.  In section \ref{sec:basic} we set out the form of the equations to be solved, without 
specialising to any particular model for the fluids themselves. In section \ref{sec:normal} we estimate diffusion timescales assuming 
that the neutron star matter is normal (i.e. not superfluid), essentially recovering part of the analysis of GR92. In section \ref{sec:sfmatter} 
we repeat the calculation assuming neutron and proton superfluidity. Finally, in section \ref{sec:dac} we summarise our findings and discuss 
their implications for the magnetar model.

%%%%%%%%%%%%%%%%%%%%%%%%%%%%%%%%%%%%%%%%%%%%%%%%%%%%%%%%%%%%%%%%%%%%%%%%%%%%%%%%%

\section{Basic model}

\label{sec:basic}

\subsection{Ambipolar diffusion model and hydromagnetic equilibrium}

\label{sec:model}

Our hydrodynamical model for the neutron star core is a magnetised three-fluid system consisting of neutrons, protons and electrons (labelled
by the index $\rx= \{\rn,\rp,\re \}$), allowing for relative motion with velocities $\vec{v}_\rx$. 
Besides the proton-electron electromagnetic coupling, our model accounts for fluid coupling through binary particle scattering or 
vortex mutual friction (depending on whether some fluids are in a superfluid state or not). In addition, our model is `transfusive' 
in the sense that it allows for particle conversion through reactions. 

Naturally, the model also includes several simplifications. 
In particular, in writing down our superfluid hydrodynamical equations of motion, we will neglect the effect of entrainment between neutrons and 
protons, which is present when both species are superfluid.  This approximation is justified in view of the order-of-magnitude precision 
of our final results\footnote{Note however that by invoking vortex-electron scattering as a possible mutual friction mechanism 
we are implicitly assuming the existence of entrainment on the mesoscopic scale of individual vortices, as it is entrainment that causes the 
protons to magnetise the neutron vortices.}.

We also need to clarify to what extent we account for the effects of the expected type II proton superconductivity in the magnetohydrodynamics
of the outer neutron star core (for a recent discussion see \citet{supercon}). A key effect we do include in our analysis is the 
superconducting magnetic force appearing in the equations of motion in the place of the familiar Lorentz force, due to the presence of the 
quantised proton fluxtubes. An effect we do {\it not} include is the participation of the fluxtubes in the mechanics of 
mutual friction coupling between the fluids. This is an important omission in our model given that the dynamics of proton fluxtubes and 
their interaction with other stellar components (e.g. the neutron vortices) may play a central role in magnetic field evolution 
(see e.g. \citet{ruderman}). The magnetic field properties (equilibrium, secular evolution) in a `full' superconducting model 
will be discussed elsewhere.    

The remaining simplifications adopted in our model have to do with the particular nature of ambipolar diffusion. By definition, this
is a slow process (i.e. it acts on timescales much longer than any dynamical timescale) with the star evolving through a series of 
`snapshots', each one of them representing a hydromagnetic quasi-equilibrium. In turn, and given that the magnetic field and particle reactions
have a small overall impact on the bulk stellar structure, we can assume that at any given time the system represents a small 
deviation from a stationary, chemically equilibrated, non-magnetic star in uniform rotation.

At this point one has to be careful with the precise notion of hydromagnetic equilibrium. In the case of a single fluid star, this notion is 
unambiguous: it refers to a state where the fluid motion is that of rigid-body rotation and where there is a balance between the 
magnetic stresses, the fluid pressure and the centrifugal force (if rotation is significant).  
This force balance is the basis of all existing calculations on hydromagnetic equilibria 
(for recent examples, see \citet{haskell08,ciolfi09,sam09}).

However, in more realistic multi-component stellar models,  transfusion (particle reactions) can occur, and the nature of 
the hydromagnetic equilibrium may change due to the ability of particle reactions in driving fluid flow and vice-versa 
(this will become apparent by inspection of the mass continuity equations (\ref{contis}) below). As will be discussed in detail, 
these flows themselves create frictional forces, from the drag of one species against another or (in the superfluid case) from the mechanism 
known as mutual friction. If the rate of particle scattering is high enough, the frictional forces connected with these diffusive flows 
might overcome the forces generated by pressure gradients and play a dominant role in balancing the magnetic stresses. 
 
We are therefore led to make a distinction between two sorts of slowly evolving equilibria: 
\emph{friction-dominated} equilibrium and \emph{transfusion-dominated} equilibrium.  
Roughly speaking, the designation is as follows.  
At high temperatures, the rate of particle reactions is high, which tends to eliminate  chemical potential imbalances 
and the corresponding pressure gradients, while the rate of particle scattering is also high, so the drag force induced by
the fluid flow is strong.
In this high temperature regime,  the key force balance is  between magnetic forces and the frictional drag forces;  
hence the designation friction-dominated equilibrium.  
Conversely, at low temperatures, the rate of particle reactions is low, so that chemical potential perturbations and the corresponding 
pressure gradients persist and are therefore important, while the rate of particle scattering is also low, so the  drag force 
associated with the diffusive flow is weak. In this low temperature regime the  key force balance is between 
the magnetic forces and the pressure gradients connected with the chemical potential imbalances;  hence the designation 
transfusion-dominated equilibrium. In our discussion below we will encounter both types of quasi-equilibria. 

Of course, no previous studies have calculated such multi-component equilibria.  The closest things that exist are the multifluid but 
non-magnetic equilibria  of \citet{prix04}, who computed stationary equilibria of rotating two-fluid stars.  
The two fluids were both rigidly rotating, and both particle reactions and particle scattering were `switched off', so as to allow solutions 
that were both stationary and did not involve frictional forces, only pressure gradients, rotation and gravity.
We will refer to such simplified equilibria (magnetised or not) as `ideal' equilibria.  
They are of significance in as much as they are the sorts of equilibria that have appeared previously in the literature, and can be 
relatively easily calculated.  Construction of an ideal  magnetic equilibrium would be a useful first step to calculating the more physical 
equilibria.
Note that  the ideal equilibrium is relatively close to the transfusion-dominated equilibrium described above, in the sense that in both, 
it is pressure forces rather than  frictional forces that are key in maintaining equilibrium.

%%%%%%%%%%%%%%%%%%%%%%%%%%%%%%%%%%%%%%%%%%%%%%%%%%

\subsection{Basic formalism}

\label{sec:formalism}

We can now assemble the equations that describe our system. According to the model defined above, when we formulate the relevant 
hydromagnetic equations we can (i) ignore time derivatives (effectively filtering out wave phenomena) and (ii) work at leading order 
with respect to the deviation from a non-magnetic background.  These same approximations were also used by GR92.

First we have the Euler equations, on a per-volume basis, 
referred to a frame rotating with the star. We write down one equation to describe the neutrons, and one to describe 
the combined proton-electron charged fluid. We will take as our background solution a rigidly rotating non-magnetic star.  
Using standard cylindrical coordinates $\{\varpi,z,\varphi\}$ with the angular velocity taking the form $\Omega \hat{z}$, 
where the hat indicates a unit vector, we have:
\bear
&& \rho_\rn \vec \nabla \left ( \frac{\mu_\rn}{m} +   \Phi - \frac{1}{2} \Omega^2 \varpi^2 \right )  = 0
\label{bgeulern}
\\ 
&&\rho_\rp \vec \nabla \left ( \frac{\mu_\rp}{m} + \frac{\mu_\re}{m} +  \Phi - \frac{1}{2} \Omega^2 \varpi^2 \right )
= 0
\label{bgeulerp}
\eear
where $\rho_\rx $ are the mass densities (with $n_\rx = \rho_\rx /m_\rx$ the corresponding number densities), $\mu_\rx$ are the chemical 
potentials and $\Phi$ is the gravitational potential. In the proton-electron equation we used charge neutrality 
$n_\rp = n_\re$ and neglected the electron mass (hereafter we use $m_\rp = m_\rn =m$).

The perturbed hydromagnetic system is described by the pair of Euler equations (adopting the so-called Cowling approximation 
which entails $\delta \Phi=0$),
\bear 
&& \rho_\rn \left [  2\vec \Omega \times \vec v_{\rm n}  + \vec \nabla \left ( \frac{\delta\mu_\rn}{m} \right ) \right ]
=\vec F_{\rm cpl} 
\label{eq:euler_n}
\\
\nonumber \\
&& \rho_p \left [ 2\vec \Omega \times \vec v_{\rm p}  +  \vec \nabla  \left ( \frac{\delta\mu_\rp  + \delta \mu_\re}{m} \right )
\right ] =  \vec F_{\rm mag} -  \vec F_{\rm cpl}  
\label{eq:euler_c}
\eear
where $\vec F_{\rm cpl}$ and  $\vec F_{\rm mag}$ are, respectively, the coupling and magnetic forces per unit volume. Their particular
form depends on the state of matter (normal or superfluid) and will be specified later.

The Euler equations are supplemented by the mass continuity equations
\be
{\vec \nabla} \cdot ( n_\rx \vec{v}_\rx ) = \Gamma_\rx
\label{contis} 
\ee
which feature the particle reaction rates\footnote{These rates also appear in the Euler equations, in the form of `rocket' terms
$\Gamma_\rx \vec{v}_\rx $ (see for example \citet{prix04}). However these are higher order terms in our perturbative scheme.}  
$\Gamma_\rx$. Baryon and charge conservation dictate that $\Gamma_\rn = -\Gamma_\rp$
and $\Gamma_\rp = \Gamma_\re$. These properties are indeed satisfied by beta decay in neutron star cores, typically operating through the 
modified Urca reactions. The background system, as described by eqns.~(\ref{bgeulern}) and (\ref{bgeulerp}),
is characterised by chemical equilibrium, i.e. $\mu_\rn = \mu_\rp + \mu_\re$. This is no longer true in the magnetically perturbed 
system. The departure from chemical equilibrium is quantified by the `chemical imbalance' parameter:
\be
\beta = \delta \mut_\rp + \delta \mut_\re - \delta \mut_\rn, \qquad \dmut_\rx = \frac{\delta \mu_\rx}{m}
\ee
For a small $\beta$ we can write
\be
\Gamma_\rn = m \lambda \beta
\ee
where the coefficient $\lambda$ is a function of density and temperature.

Note that for the ideal equilibrium defined earlier, the Coriolis terms in the Euler equations would be zero (all fluids rotate rigidly), 
the coupling force $\vec{F}_{\rm cpl}$ would be set to zero, and particle reactions eliminated by setting $\lambda = 0$.

Returning to the Euler equations (\ref{eq:euler_n}) and (\ref{eq:euler_c}) we can combine them 
and produce an equivalent (but slightly more useful) set of equations. Taking the difference between the two 
equations and manipulating slightly gives
\be
 2\Om (\hat{z} \times  \vec{w} )  + \vec \nabla \beta + \frac{1}{\xp \rho_\rn} \vec{F}_{\rm cpl} 
= \frac{1}{\rho_\rp} \vec{F}_{\rm mag} 
\label{diffeq}
\ee 
where $\xp = \rho_\rp/\rho $ is the proton fraction ($\rho = \rho_\rn +\rho_\rp$ is the total density) and
\be
\vec{w} = \vec{v}_\rp - \vec{v}_\rn 
\ee
We will refer to (\ref{diffeq}) as the `difference Euler equation'.

Taking instead the sum of the two equations gives
\be
2\Om (\hat{z} \times \vec{u} )  +  \vec \nabla \dmut_\rn + \xp \vec \nabla \beta  =  \frac{1}{\rho} \vec{F}_{\rm mag}
\label{aveq}
\ee 
where $\vec{u} = \vec{v}_\rn  + \xp \vec{w}$ 
is the neutron-proton average velocity, weighted by density. We will refer to (\ref{aveq}) as the `total Euler equation'.  

By combining the continuity equations in a similar way we get the set,
\bear
&& \vec \nabla \cdot ( \rho \vec{u} ) = 0
\label{avecont}
\\
&&  {\vec \nabla} \cdot ( n_\rp \vec{w}) +  n_\rn {\vec \nabla} \left (\frac{n_\rp}{n_\rn} \right ) \cdot \vec{v}_\rn 
= -\left (1 +\frac{n_\rp}{n_\rn} \right ) m \lambda \beta 
\label{diffcont}
\eear
Strictly speaking, the two terms on the left hand side of equation (\ref{diffcont}) could have comparable magnitudes, 
unless we invoke a weak stratification (in the sense that $\xp$ and therefore also $n_\rp/ n_\rn$ vary only slightly over length scales of 
interest) which would allow us to neglect the second term. However, and despite its common use by many authors
(including GR92), this is not a well justified approximation in realistic neutron star models.   Perhaps  a better reason for 
neglecting the second term in (\ref{diffcont}) would be to use (\ref{avecont}) which suggests a meridional neutron velocity 
component $ \hat{\varphi} \times \vec v_\rn \sim x_\rp (\hat{\varphi} \times \vec w)$. 
Following this reasoning we will neglect the second term in (\ref{diffcont}), but it should be pointed out that {\em none} of our 
(order-of-magnitude precision) results and conclusions would change had we kept this term, as long as $v_\rn \lesssim w$. Hence, we write
\be
{\vec \nabla} \cdot ( n_\rp \vec{w})  = - \tilde\lambda \beta 
\label{eq:c_cont}
\ee
where we have defined
\be 
\tilde\lambda =  \left (1 +\frac{n_\rp}{n_\rn} \right ) m \lambda  . 
\ee

Before proceeding to our discussion of ambipolar diffussion we can make some non-trivial observations based on the `difference'
and `total' equations (\ref{diffeq}) and (\ref{aveq}) in the limit of an ideal hydromagnetic equilibrium 
(i.e. $\vec v_\rx = 0,~\lambda =0,~\vec{F}_{\rm cpl} =0 $). Firstly, from eqn. (\ref{diffeq}), we see that a star 
endowed with a magnetic field {\it cannot} be in chemical equilibrium, as it is the chemical imbalance $\beta$ that 
provides the `pressure' gradient $\vec \nabla \beta$ necessary to balance $\vec F_{\rm mag} / \rho_{\rm p}$, the magnetic force per 
unit mass of charged  plasma. At the same time, from eqn. (\ref{aveq}) we see that only in the special case when $x_\rp$ is a function of 
$\beta$ is $\vec F_{\rm mag} / \rho$, the magnetic force per unit total mass, equal to the gradient of a scalar. This is reminiscent of the case 
of a single-fluid barotropic star, where the notions of stratification and chemical equilibrium are not relevant, and $\vec F_{\rm mag} / \rho$ 
is always balanced by the gradient of a scalar (e.g. \citet{haskell08}). Finally, from either Euler equation it follows that an axisymmetric 
system has a vanishing azimuthal magnetic component $F^\varphi_{\rm mag} =0$.

%%%%%%%%%%%%%%%%%%%%%%%%%%%%%%%%%%%%%%%%%%%%%%%%%%%%%%%%%%%%%%%%%%%%%%%%%%%

\subsection{General strategy}
\label{sec:strategy}

Our strategy will now be to look for solutions to the above set of equations, where we regard the magnetic field $\vec B$ and 
therefore also $\vec F_{\rm mag}$ as given, so that the magnetic force on the right hand sides of eqns. (\ref{diffeq}) and (\ref{aveq}) 
play the role of source terms, driving departures from chemical equilibrium and ambipolar diffusion of the charged particles with respect to 
the neutrons.  We will assume that the magnetic field itself is axisymmetric, and that the subsequent evolution preserves this axisymmetry.  
This analysis is very much in the spirit of GR92--we only seek to make rough estimates of timescales, so will not attempt to construct 
fully self-consistent solutions, neither for the background nor for the perturbations.

The departure from chemical equilibrium is given by the chemical imbalance $\beta$.  We will use the proton-neutron velocity difference 
$\vec w$ as our diagnostic of ambipolar diffusion, and refer to $\vec w$ as the ambipolar diffusion velocity.  Of course, ambipolar diffusion 
refers to the combined motion of the charged species (protons plus electrons) relative to the neutrons, so for the use of the term 
`ambipolar diffusion'  to be strictly accurate, it is necessary that the electron-proton velocity difference 
$\vec w_\rpe = \vec v_\rp - \vec v_\re$ be smaller than the proton-neutron velocity difference:
 \be
w \gg w_\rpe
\label{cond}
\ee   
We will check if this inequality is satisfied in our calculations below.  Note that when it is not satisfied, we would still expect 
the charged particle flow to generate magnetic field evolution; it is simply that it is no longer appropriate to describe it as 
classical ambipolar diffusion.

In the calculations that follow, we will use the perturbation equations to make rough estimates of the ambipolar diffusion velocity 
$w = |\vec w|$, which can then be immediately converted into timescales ($\tau \equiv L / w$, where $L$ is a characteristic length scale).
Within the order-of-magnitude precision of this calculation the timescale $\tau$ is essentially equal to the evolution timescale 
$\sim L/v_\rp$ derived by the magnetic induction equation (see, for example, GR92).
We will make repeated use of eqn. (\ref{diffeq}), the difference Euler equation. As noted above, the magnetic force on the right hand of 
this equation is to be regarded as the driving term, sourcing ambipolar diffusion.  The terms on the left hand side all depend linearly on $w$.  
For the Coriolis term this dependence is explicit.  For the coupling term this dependence is a function of the assumed state of matter (normal 
verses superfluid), as will be described below.  The term $\vec \nabla \beta$ represents the difference in pressure forces on the neutron and 
charged fluids, and can be related to the ambipolar diffusion velocity via eqn. (\ref{eq:c_cont}).   
In the next two sections we explore the ambipolar timescales for normal and superfluid neutrons, accounting for the relative importance of 
the different terms as a function of temperature.

%%%%%%%%%%%%%%%%%%%%%%%%%%%%%%%%%%%%%%%%%%%%%%%%%%%%%%%%%%%%%%%%

\section{Ambipolar diffusion: normal matter}
\label{sec:normal}

Before tackling the superfluid problem, we will look at the case of normal matter. We will assume that the particle 
reactions are those of modified Urca, with an efficiency factor \citep{sawyer}
\be
\lambda = 2.5 \times 10^{27}\,  \rho_{14}^{2/3} T^6_8 \, ~\mbox{erg}^{-1} \mbox{cm}^{-3}\mbox{s}^{-1}
\ee 
where $T_8 = T/(10^8$ K) and $\rho_{14} = \rho/(10^{14}$ g/cm$^3$).

The magnetic force can be identified with the usual Lorentz force (per unit volume):
\be
\vec{F}_{\rm mag}  = \frac{1}{4\pi} (\vec \nabla \times \vec B )\times \vec B  
\label{lorentz}
\ee
The coupling forces arise from binary particle collisions, so that the force is a function of the particle collision 
frequencies ${\cal D}_{\rm xy}$:
\be
\vec F_{\rm cpl} =  \rho_{\rm p} {\cal D}_{\rm pn} \vec w  + \rho_{\rm e} {\cal D}_{\rm en} \vec w_{\rm en} . 
\ee
where $\vec w = \vec v_{\rm p} - \vec v_{\rm n}$ as defined previously, and $\vec w_{\rm en} = \vec v_{\rm e} - \vec v_{\rm n}$.  
These collision frequencies have been calculated by \citet{ys90}, who found that the neutron-proton scattering (mediated by 
the strong nuclear force) was much more frequent  than the neutron-electron scattering (mediated by the weak nuclear force), so 
we will include only the former in our coupling force:
\be
\vec F_{\rm cpl} \approx \rho_{\rm p} {\cal D}_{\rm pn} \vec w
\label{Fdrag}
\ee
where 
\be
D_\rpn  \approx 6.6 \times 10^{16}\, T^2_8 \rho_{14}^{-1/3}\, {\rm s}^{-1} .
\ee
The difference and total Euler equations then become
\bear
&& 2\Om ( \hat{z} \times \vec w) + \cD \vec{w} + \vec \nabla \beta  = \frac{1}{\rho_\rp} \vec{F}_{\rm mag}
\label{diff_norm}
\\
\nonumber \\
&& 2\Om  ( \hat{z} \times \vec u  ) + \vec \nabla \dmut_\rn  + \xp \vec \nabla \beta =  \frac{1}{\rho} \vec{F}_{\rm mag}
\label{ave_norm}
\eear
where $ \cD = (\rho/\rho_\rn) D_\rpn$.

As we will now show, rotation has a minimal impact on ambipolar drift motion. Indeed, we can see that 
the ratio of the Coriolis force to the drag force is $\sim 2\Om /\cD \ll 1$. It turns out that the same ratio 
controls the magnitude of the azimuthal component of the ambipolar drift. To see this, note that from 
eqns. (\ref{diff_norm}) and (\ref{ave_norm}) our assumption of axisymmetry implies an azimuthal component
$F^\varphi_{\rm mag} = \cD w^\varphi + 2\Om w^\varpi $ and $ F_{\rm mag}^\varphi = 2\Om ( v^\varpi_\rn + x_\rp w^\varpi )$,
respectively. Equating these
\be
  \frac{w^\varphi}{w^\varpi} = \frac{2\Om}{\cD} \left ( \frac{v^\varpi_\rn}{w^\varpi} + x_\rp -1 \right)
  \quad \Rightarrow \quad 
  \frac{w^\varphi}{w^\varpi} \sim \frac{2\Om}{\cD} \ll 1 
  \label{wratio1}
\ee 
which implies a dominantly {\it poloidal} ambipolar flow (this conclusion assumes
that $v_\rn^\varpi$ is not exceedingly larger than $w^\varpi$, which is a reasonable expectation).
Note that if we were to consider a non-rotating star then the same argument would have led to vanishing azimuthal components
$F^\varphi_{\rm mag} = w^\varphi =0$.

We now wish to use (\ref{diff_norm}) to make an estimate of the ambipolar diffusion velocity. At this point GR92 carried out a 
decomposition on eqn. (\ref{diff_norm}) itself, splitting the equation into its solenoidal (zero divergence) and irrotational (zero curl) 
parts. This decomposition has the attractive feature that the $\vec \nabla \beta$ term is purely irrotational, so the solenoidal projection 
of the equation represents a balance purely between the drag and Lorentz forces.  However, there is no simple way of decomposing the Lorentz 
force itself into its solenoidal and irrotational parts, so we will instead make use of  a slightly different decomposition.  
We will decompose at the level of the ambipolar velocity $\vec w$, splitting it into what we will loosely term its `solenoidal' and 
`non-solenoidal' parts
\be
  \vec w = \ws + \wns  
  \label{w_deco}
\ee
where $\ws$ and $\wns$ are defined to satisfy
\be
  {\vec \nabla} \cdot ( n_\rp \ws)  = 0 
  \hspace{10mm}
  {\vec \nabla} \cdot ( n_\rp \wns)  = -\tilde\lambda \beta 
  \label{conti_deco}
\ee
With these formal definitions, the non-solenoidal part of the velocity field is that part of $\vec w$ which is sensitive to 
particle transfusion, 
while the solenoidal part is not associated with perturbations in the chemical potentials/pressure.  
When inserted into (\ref{diff_norm}), and dropping the Coriolis force, this gives
\be
  \cD (\ws + \wns)
  - \vec \nabla  \left[ \frac{1}{\tilde\lambda} \vec\nabla \cdot(n_\rp \wns) \right]
  = \frac{1}{\rho_\rp} \vec{F}_{\rm mag}
  \label{diff_deco}
\ee
To make progress we can exploit that fact that both the drag coefficient $\cal D$ and the reaction parameter $\tilde\lambda \approx m\lambda$ 
are temperature dependent, allowing us to compare the relative sizes of the drag and pressure gradient forces acting upon the non-solenoidal 
flow: 
\be
  \alpha = \frac{{\cal D} \wnss}{\nabla \beta} \sim \frac{D_\rpn \lambda m^2  L^2}{\rho_\rp} 
  \approx 5 \times  10^{-5}\,\frac{L_6^2 T^8_8}{ \rho_{14}^{2/3} x_{01}}
  \label{trans}
\ee
where we used the normalisations $L_6 = L/(10^6\,\mbox{cm}), ~x_{01} = \xp/0.1$. 
This ratio becomes unity at a temperature
\be
  \alpha =1 \quad \Rightarrow \quad T_\alpha \sim 4 \times 10^{8}\, \frac{\rho_{14}^{1/12} x_{01}^{1/8}}{L_6^{1/4}} \,~ \mbox{K} 
  \label{Ta}
\ee  

Given the steep  temperature dependence of the ratio $\alpha$, it makes sense to consider the low and high temperature cases separately.  
If we begin with the $T > T_\alpha$ regime, from eqn. (\ref{trans}) we see that we can neglect the $\beta$ `pressure' gradient term  in 
(\ref{diff_deco}) to give
\be
\cD (\ws + \wns) \approx \frac{1}{\rho_\rp} \vec{F}_{\rm mag}
\label{high_T_forces}
\ee
We see that at high temperatures, the drag forces dominate the pressure gradients and ambipolar diffusion represents a balance between 
the drag force and the Lorentz force. We therefore see that for $ T > T_\alpha$ we are in the friction-dominated regime of 
section~\ref{sec:model}, or, equivalently, the star is in a friction-dominated equilibrium. 
In this regime the decomposition (\ref{w_deco}) and (\ref{conti_deco}) is no
longer useful and we can estimate a common diffusion timescale 
\be
\tau \sim \frac{L}{w}  \sim
\frac{4\pi \rho_{\rm p} L^2 D_{\rm pn}}{B^2} 
\approx 3 \times 10^5\, x_{01} \rho_{14}^{2/3} L_6^2 \frac{T_8^2}{B_{15}^{2}} {\, \rm yr}.
\label{tauhot}
\ee
where $B_{15} = B/(10^{15}$ G). 

For $T < T_\alpha$, eqn. (\ref{trans}) shows we can instead neglect the effect of drag on the 
non-solenoidal flow, so that eqn. (\ref{diff_deco}) becomes
\be
\cD \ws - \vec \nabla  \left[ \frac{1}{\tilde\lambda} \vec\nabla \cdot(n_\rp \wns) \right]
 \approx \frac{1}{\rho_\rp} \vec{F}_{\rm mag}
\label{diff_deco1}
\ee
We now need to be careful in extracting a diffusion timescale. In fact we can calculate two distinct timescales, by following GR92 and  
identifying that  part of the Lorentz force that represents a departure of $\vec{F}_{\rm mag}/\rho_\rp$ from a perfect gradient:
\be
\vec \cF_{\rm mag} = \vec{F}_{\rm mag} - \rho_\rp \vec \nabla \beta
\ee
Recall that in the single fluid case, the force $\vec{F}_{\rm mag} / \rho$  was purely irrotational.  Unfortunately, as discussed in section 
\ref{sec:formalism}, there does not yet exist a calculation of the magnetic field structure of a multifluid star, so we cannot be sure  
how $\vec \cF_{\rm mag}$ compares in magnitude with the full Lorentz force. For a generic  initial stellar field, it is conceivable 
that $|\vec \cF_{\rm mag}| \sim |\vec{F}_{\rm mag}|$. Another possibility (once the system enters the $T<T_\alpha$ regime) is that the flow
$w^\rs$, and therefore $\vec \cF_{\rm mag}$,  is quenched as a result of short-timescale, dynamical evolution; if viable, this process
cannot be described by our model. Given our ignorance on this point, all we can say is that $|\vec \cF_{\rm mag}| \lesssim |\vec{F}_{\rm mag}|$ 
and we will parameterise accordingly in the equations that follow.

With this in mind,  one timescale is associated with the non-solenoidal flow maintained by the non-zero $\beta$ which is responsible for balancing 
the magnetic force. This is described by  
\be
- \vec \nabla   \left[ \frac{1}{\tilde\lambda} \vec\nabla \cdot(n_\rp \wns) \right]
 \approx \frac{1}{\rho_\rp} \vec{F}_{\rm mag}
 \label{forces_ns}
\ee
which leads to an ambipolar diffusion timescale
\be
\tau^{\rm ns} \sim \frac{L}{\wnss} \sim 
\frac{4\pi\rho_\rp^2}{m^2 \lambda B^2} \approx
6 \times 10^9\, \frac{x_{01}^2 \rho_{14}^{4/3}}{T^{6}_8 B_{15}^{2}} {\, \rm yr}.
\label{taucold1}
\ee
The second timescale is associated with the solenoidal flow sourced by $\vec \cF_{\rm mag}$,
\be
\cD \ws = \frac{1}{\rho_\rp} \vec \cF_{\rm mag}
\label{diff_sol}
\ee
Given that $|\vec \cF_{\rm mag}| \lesssim |\vec{F}_{\rm mag}|$ we obtain a lower bound on the associated diffusion timescale:
\be
\tau^{\rm s} \sim \frac{L}{\wss} \sim \frac{ \rho_\rp \cD L}{\cF_{\rm mag}} \gtrsim
3 \times 10^5\, x_{01} \rho_{14}^{2/3} L_6^2 \frac{T_8^2}{B_{15}^2} {\, \rm yr}.
\label{taucold2}
\ee
The timescales %(\ref{tauhot}) and (\ref{taucold1}) 
given above are in agreement with the ones obtained by
GR92, aside from our solenoidal timescale (\ref{taucold2}) being a lower bound rather than an estimate (GR92 implicitly assumed 
$|\vec \cF_{\rm mag}| \sim |\vec{F}_{\rm mag}|$). This level of agreement is not surprising given that, so far, our analysis closely followed 
the one in GR92.

The following scenario then suggests itself: after the birth of a neutron star, once all the fast dynamical perturbations not considered here 
have died away, the hot ($T > T_\alpha$) star cools, proceeding through a series of friction-dominated equilibria as described by 
eqn. (\ref{high_T_forces}).  When the star cools below $T_\alpha$, the evolution can then be dominated by the solenoidal flow of 
eqn. (\ref{diff_sol}), provided that $\ws$ is not suppressed on a short dynamical timescale.   
%which would act to eliminate  any solenoidal part of the magnetic field that the star has retained from birth.  
Only on timescales longer than that of eqn. (\ref{taucold2}), after which the solenoidal flow is likely to die away, 
will the evolution proceed slowly through a series of transfusion-dominated equilibria as given by eqn. (\ref{taucold1}). 
The current magnetar activity might then be the result either of the combined solenoidal/non-solenoidal flow of eqn. (\ref{high_T_forces}) 
or the solenoidal flow of eqn. (\ref{diff_sol}). In either case,  it follows that if the initial stellar magnetic 
field does indeed contain a significant $\vec \cF_{\rm mag}$ component, the currently observed magnetars may well be in a 
regime of friction-dominated equilibrium, in which case all previous calculations of their equilibrium structure might be seriously 
in error  (e.g. \citet{haskell08,ciolfi09,sam09}).

Finally, we can check if this motion represents ambipolar diffusion, in the sense of eqn. (\ref{cond}), 
i.e. we need to verify that the proton-electron relative speed $w_\rpe$ is much smaller than the ambipolar drift velocity $w$.  
The speed $w_\rpe$ can be easily estimated from the Amp\`ere law to be
\be
 w_\rpe  \sim 8 \times 10^{-10} \frac{B_{15}}{L_6 x_{01} \rho_{14}} {\, \rm cm \, s^{-1}}
\label{wpe}
\ee
while, for the high temperature flow of eqn. (\ref{high_T_forces}), or for the low-temperature solenoidal flow of 
eqn. (\ref{diff_sol}), we have
\be
w \sim 10^{-7}  \frac{B_{15}^2}{x_{01} L_6 T_8^2 \rho_{14}^{2/3}} {\, \rm cm \, s^{-1}}
\ee
We therefore see that condition (\ref{cond}) is satisfied for stars with temperatures of order of a few times $10^8$.  
However, for the non-solenoidal mode of eqn. (\ref{forces_ns}), we instead have
\be
\wnss \sim 6 \times 10^{-11} \frac{B_{15}^2 L_6 T_8^6}{x_{01}^2 \rho_{14}^{4/3}} {\, \rm cm \, s^{-1}}
\ee
which is comparable to the velocity estimate of  eqn. (\ref{wpe}).  We therefore see that in this case the rapid motion of the protons with respect 
to the neutrons implies that this motion cannot be interpreted as  ambipolar diffusion in the traditional sense.  However, as described in 
section \ref{sec:strategy}, we would nevertheless expect the magnetic field structure to evolve on the timescales estimated. 
The conceptual difficulty associated with the violation of the condition (\ref{cond}), although potentially important for ambipolar diffusion 
in normal matter, becomes irrelevant in the presence of proton superconductivity (as we discuss in section~\ref{sec:sfhot}).

%%%%%%%%%%%%%%%%%%%%%%%%%%%%%%%%%%%%%%%%%%%%%%%%%%%%%%%%%%%%%%%%%%%%%%%%%%%

\section{Ambipolar diffusion: superfluid matter}
\label{sec:sfmatter}

In the presence of superfluidity it is clear that the preceding ambipolar diffusion model needs revision. 
Once a given particle species is in a superfluid state, the rates of physical processes where these particles participate
(e.g. binary collisions and chemical reactions) are exponentially suppressed. Hence, we need to make the following
modifications to the normal matter model:
\be
\cD_\rpn \to \cR_\rpn \cD_\rpn, \qquad \lambda \to  \cR_{\rm sf} \lambda
\label{suppress}
\ee
where $\cR_\rpn $ and $\cR_{\rm sf}$ are superfluid suppression factors. These are exponentially small numbers for 
temperatures well below the critical temperatures $T_{\rm cn}$ and $T_{\rm cp}$ for neutron and proton
superfluidity. In the $T \ll T_{\rm cn}, T_{\rm cp}$ regime, and as long as the critical temperatures are not too different, 
these factors can be approximated by the following fits to the available rigorous results \citep{haensel00,haensel01,baiko01},
\be
{\cal R}_\rpn \approx 0.01\, y^2_\rn y_\rp\, e^{-(y_\rn + y_\rp)}, 
\quad 
{\cal R}_{\rm sf} \approx 3 \times 10^{-4} y_\rp^5 e^{-2y_\rn} 
\label{Rfits}
\ee
where $y_\rn = 1.188\, T_{\rm cn}/T$ and $y_\rp = 1.764\, T_{\rm cp}/T$ (the difference in the numerical factors reflects
the different type of pairing expected in the outer core: ${}^1S_0$ pairing for the protons  and ${}^3P_2$ pairing for 
the neutrons, see, for example, \citet{sauls}). Note that in general  $\cR_\rnp \neq \cR_\rsf$.

As a result of the combined superfluid suppression in particle reactions and collisions, the notion of a friction-dominated 
equilibrium (section~\ref{sec:model}) becomes irrelevant; this is demonstrated below in detail. Thus, a superfluid star is 
expected to be close to an ideal hydromagnetic equilibrium (i.e. in the transfusion-dominated regime) for any temperature below the 
critical ones.

Besides the suppression described by (\ref{suppress}), superfluidity also adds a new coupling mechanism between the fluids through vortex 
mutual friction. Mutual friction refers to the force mediated by the neutron vortices (which are responsible for the star's bulk 
rotation) as a result of their interaction with the charged particles and the magnetic field. The standard expression for the resulting 
coupling force (per unit volume, acting on the neutrons) is given by \citep{HV,na06},
\be
\vec{F}_{\rm mf} = -2 \rho_\rn   \Omega \left [ \cBp ( \hat{z} \times \vec{w} ) + 
\cB \left \{ (\hat{z} \cdot \vec{w} ) \hat{z} - \vec{w} \right \}  \right ]
\label{mfforce}
\ee
where
\be
\cB = \frac{\cR}{1+\cR^2}, \qquad \cBp = \frac{\cR^2}{1+\cR^2}
\ee
The dimensionless coefficient $\cR$ is associated with the effective drag force experienced by a moving vortex.
For `standard' mutual friction, i.e. scattering of electrons off the vortex's magnetic field,
the drag value is estimated to be \citep{als88,na06}
\be
\cR \approx 4\times 10^{-4} \quad \to \quad \cBp \approx \cB^2, \quad \cB \approx \cR \ll 1
\label{weak}
\ee
In this case the coupling between the neutron fluid and the charged particles can be considered
{\it weak}. The opposite limit of {\it strong} mutual friction corresponds to
\be
\cR \gg 1 \quad \to \quad  \cBp \approx 1, \quad \cB \approx \frac{1}{\cR}
\label{strong}
\ee
Strong drag could emerge as a result of interactions between vortices and fluxtubes \citep{sauls,ruderman,link03}. 
The extreme case of perfect {\it vortex  pinning} corresponds to $\cBp \to 1$ and $\cB \to 0$.  We will {\it not} consider the case of 
perfect pinning in this paper, as it would require a model which incorporates the detailed dynamics of neutron vortices and proton fluxtubes.
These effects are beyond the scope of this paper and will be addressed elsewhere.

The total coupling force is now the sum of the drag and mutual friction forces:
\be
\vec F_{\rm cpl} = \vec F_{\rm drag} + \vec F_{\rm mf}
\ee
where the drag force takes on the same functional form as in the normal fluid case, aside from the presence of the 
suppression factor $\cal R_{\rm pn}$:
\be
\label{eq:sf_drag}
\vec F_{\rm drag} = \rho_{\rm p} {\cal R}_{\rm pn} {\cal D}_{\rm pn} \vec w
\ee
The drag force is strongly suppressed at low temperatures, so that in a sufficiently cold star the mutual friction force will 
dominate the coupling. However, it is possible that for higher temperatures (but still below the critical ones) the drag force could 
dominate the mutual friction, despite the suppression of particle collisions. We can quantify this argument by means of  the ratio 
\be
f = \frac{F_{\rm drag}}{F_{\rm mf}} \sim \frac{\cR_\rpn D_\rpn \xp}{2 \Om \cB_{\rm max}}  
\approx  5 \times 10^{14}\, \cR_\rpn \frac{ x_{01} T^2_8}{\cB_{\rm max}} \left ( \frac{P}{1\,\mbox{s}} \right )
\label{fratio}
\ee
where $\cB_{\rm max} = {\rm max} [\cB,\cBp]$. Depending on whether $f \ll 1$ or $f \gg 1 $ we will refer to the 
`cold superfluid' and `hot superfluid' regimes, respectively. The transition between these two regimes occurs at a 
temperature $T_f$ at which $f \sim 1$ and, due to the exponentially varying $\cR_\rnp$, it is very rapid.
We can estimate $T_f$ by considering the simple case where $T_{\rm cn} = T_{\rm cp} = T_\rc$ and adopting
the weak mutual friction limit (\ref{weak}). Using a typical value $T_\rc = 5 \times 10^9\, \mbox{K}$ in (\ref{Rfits}) 
we find that the interval (say) $ 0.1 < f < 10 $ corresponds to
\bear
&& T_f \approx  2.7 - 2.9 \times 10^8\,\mbox{K}, \qquad P = 10\, \mbox{s}
\label{Tfnum}
\\
&&  T_f \approx  2.9 - 3.2 \times 10^8\,\mbox{K}, \qquad P = 0.1\, \mbox{s}
\eear
illustrating the steep variation of $f$ with temperature, and the weak dependence of $T_f$ on spin period.
These estimates for $T_f$ can be slightly reduced if $T_\rc$ is decreased or $T_{\rm cn} < T_{\rm cp}$
(as predicted by some pairing models, see discussion in \citet{viscous}).

The transition from `hot' to `cold' superfluid hydrodynamics is therefore potentially relevant for relatively
hot neutron stars. The existing state-of-art cooling 
calculations \citep{pmg09}  suggest that magnetar interior temperatures are expected to takes values of a few times
$10^8\, \mbox{K}$, (at the observed spindown ages $10^3-10^4\,\mbox{yr}$). 
Thus, magnetars could be found in {\it either}  superfluid regime and our modelling should be able to 
account for both possibilities.  We will therefore look at each case in turn in the following two subsections.

A final important effect of superfluidity (and in particular of type II proton superconductivity) is related to the magnetic
force $\vec F_{\rm mag}$. This force is {\it not} given by the Lorentz force (\ref{lorentz}) but, instead, is related to the
smooth-averaged self-tension of the proton fluxtube array. The explicit form of the superconducting magnetic force 
is given in \citet{supercon}.  For the purposes of this paper it is sufficient to use the
rough estimate
\be
\label{Fsuper}
 F_{\rm mag}  \sim \frac{H_{\rm c1} B}{4\pi L}
\ee
where $ H_{\rm c1} \approx 10^{15}\,\mbox{G}$ is the critical magnetic field associated with the self-energy of a single fluxtube \citep{tilley}, 
whose density dependence we will neglect.

%%%%%%%%%%%%%%%%%%%%%%%%%%%%%%%%%%%%%%%%%%%%%%%%%%%%%%%%%

\subsection{The `cold superfluid'  regime}
\label{sec:sfcold}

We first discuss the regime  $T < T_f $ where `cold' superfluid hydrodynamics applies, i.e. $ f \ll 1$, so that the coupling force can be 
approximated as being entirely due to mutual friction. In order to account for vortex mutual friction we obviously need to retain a non-zero 
rotation $\Om$ in our equations.  Also, we will continue to assume an axisymmetric system.  

To begin with, it is convenient to decompose the various velocities into poloidal and toroidal components
using standard cylindrical coordinates. For example,
\be
\vec{w} = \vec{w}_\rP + w^\varphi \hat{\varphi}, \qquad 
\vec{w}_\rP =  w^\varpi \hat{\varpi} +  w^z \hat{z}
\ee
Inserting these in the Euler eqns. (\ref{diffeq}) and (\ref{aveq}), together with the force 
(\ref{mfforce}), we obtain the difference Euler equation 
\bear
\nonumber \\
&&\frac{2 \Om}{\xp} \cB \left (  w^\varpi \hat{\varpi} + w^\varphi \hat{\varphi} \right ) + 
\vec{\nabla} \beta
\nonumber \\
&&  + 2 \Omega \left ( 1-\frac{\cBp}{\xp} \right ) \left ( w^\varpi \hat{\varphi}  
-w^\varphi \hat{\varpi} \right )  = \frac{1}{\rho_\rp} \vec{F}_{\rm mag}
\label{sfdiffeq}
\eear
and the total Euler:
\bear
&&  2\Om \Bigg [  \left ( v_\rn^\varpi + \xp w^\varpi \right ) \hat{\varphi}
-  \left ( v_{\rn}^\varphi + \xp w^\varphi  \right ) \hat{\varpi}  \Bigg ] 
+ \vec{\nabla} \dmut_\rn  
\nonumber \\
&& +  \xp \vec{\nabla} \beta =  \frac{1}{\rho} \vec{F}_{\rm mag}
\label{sfaveq}
\eear
The toroidal and poloidal parts of the difference Euler equation are:
\bear
&& 2 \Om \Bigg [ \frac{\cB}{\xp} w^\varphi +  \left ( 1-\frac{\cBp}{\xp} \right ) w^\varpi \Bigg ]
=\frac{1}{\rho_\rp} F_{\rm mag}^\varphi
\label{sfdiff_phi}
\\
\nonumber \\
&& 2\Om \hat{\varpi} \Bigg [ \frac{\cB}{\xp}  w^\varpi - \left ( 1-\frac{\cBp}{\xp} \right ) w^\varphi \Bigg ]     
+ \vec \nabla \beta  = \frac{1}{\rho_\rp} \vec F_{\rm mag}^\rP
\label{sfdiff_pol}
\eear 
Based on these general equations we can arrive at several non-trivial results. Firstly, the $z$ component of (\ref{sfdiff_pol}) shows that 
the superfluid system does not allow the establishment of chemical equilibrium ($\beta=0$)  unless $F^z_{\rm mag}=0$. 
However, it is almost certain that a generic magnetic field configuration does not comply with this requirement. 
Secondly, we can show that the magnetic force is nearly poloidal, i.e. $ F^\varphi_{\rm mag}  \ll F^\rP_{\rm mag} $, 
which is consistent with the proximity of the system (see eqn.~(\ref{asf}) below) to an ideal axisymmetric hydromagnetic equilibrium  
(which has $F^\varphi_{\rm mag} =0$ exactly). 
Indeed, from (\ref{sfaveq}) and (\ref{sfdiff_pol}) we have
\be
F^\varphi_{\rm mag} = 2\Om \rho \left ( v_\rn^\varpi + \xp w^\varpi \right ), 
\qquad  F^z_{\rm mag}  \sim \frac{\rho_\rp \beta}{L}
\label{Fphi}
\ee
Given that $ F_{\rm mag}^\rP \sim  F_{\rm mag}^z $ we have
\bear
&& \frac{F^\varphi_{\rm mag}}{F^\rP_{\rm mag}} \sim \frac{2 \Om m^2 L }{\beta} w^\varpi  
\left ( 1  + \frac{v^\varpi_\rn}{x_\rp w^\varpi} \right )
\nonumber \\
&& \qquad \quad \approx  10^{-20}\, \cR_{\rm sf} \frac{ L_6^2 T^6_8}{\rho_{14}^{1/3}x_{\rm 01}} \left ( \frac{1\,\mbox{s}}{P}  \right ) \ll 1 
\label{Fratio}
\eear
where we assumed that the bracket term is of order unity.
Finally, by combining  (\ref{sfdiff_phi})  and(\ref{Fphi})  we obtain, 
for both weak and strong mutual friction,
\be
w^\varphi = \frac{1}{\cB} \Bigg ( \cBp w^\varpi + v^\varpi_\rn \Bigg ) 
\quad \Rightarrow \quad  w^\varphi \gg  w^\varpi
\label{wt1}
\ee
In other words, superfluid ambipolar diffusion is dominantly toroidal.
This is in contrast with the properties of ambipolar diffusion in a non-superfluid star, see eqn.~(\ref{wratio1}).

In order to study ambipolar diffusion within the present superfluid model we will mainly rely on the components
(\ref{sfdiff_phi}) and (\ref{sfdiff_pol}) of the difference Euler equation, as we did in the non-superfluid case. 
Moreover, we will continue to decompose the ambipolar velocity $\vec w$ into solenoidal and non-solenoidal components, 
as in eqn. (\ref{w_deco}).

Using the property (\ref{wt1}) we can approximate (\ref{sfdiff_pol}) as
\be
\hat{\varpi} \cD_{\rm sf} w^\varphi  + \vec \nabla \beta =  \frac{1}{\rho_\rp} \vec F_{\rm mag}^\rP
\label{sfmaster}
\ee
where the effective `drag' coefficient $\cD_\rsf$ takes the following form in the limits of
weak and strong mutual friction:
\be
\cD_{\rm sf}^{\rm weak} \approx -2\Om , \qquad 
\cD_{\rm sf}^{\rm strong} \approx \frac{2\Om}{\xp}
\ee
To make our strategy maximally transparent, we can eliminate $\beta$ from eqn. (\ref{sfmaster}) using the continuity equation
\be
\vec{\nabla} \cdot ( n_\rp \vec{w}^{\,\rns}_\rP) = - \cR_{\rm sf} m \tilde{\lambda} \beta
\label{sfcont1}
\ee
making the decomposition into solenoidal and non-solenoidal velocity terms explicit:
\be
\hat{\varpi} \cD_{\rm sf} (w_\varphi^{\rm s} + w_\varphi^{\rm ns} ) 
-\vec \nabla   \left[ \frac{1}{{\cal R}_{\rm sf} \tilde\lambda} \vec\nabla \cdot(n_\rp \wns_\rP) \right]
=  \frac{1}{\rho_\rp} \vec{F}_{\rm mag}^\rP
\label{sfdiff_deco}
\ee
This equation is the superfluid analogue of eqn. (\ref{diff_deco}), and shows that the structure of the ambipolar 
diffusion problem in the superfluid case very closely mirrors that of the normal case: we have a driving term, build out of the Lorentz force, 
on the right hand side, while on the left we have terms linear in the ambipolar velocity.

We can now proceed much as in the normal case, described in section \ref{sec:normal}.  The terms on the left hand side of eqn. 
(\ref{sfdiff_deco}) are strongly temperature dependent, so we will examine the ratio of the non-solenoidal mutual friction coupling force 
and pressure gradient term:
\be
\alpha_{\rm sf} = \frac{\cD_{\rm sf} \wnss_\varphi}{\nabla \beta} 
\sim 10^{-18}\, \cR_{\rm sf} \left ( \frac{10^{-4}}{\cB} \right ) \left ( \frac{1\,\mbox{s}}{P} \right )  T^6_8 
\label{asf}
\ee
where we used $\wnss_\varphi \sim -\xp \wnss_\varpi /\cB$ for weak mutual friction (the strong mutual friction result differs
only by a factor $\xp$). 
Hence $\alpha_{\rm sf} \ll 1$ for any relevant temperature and so we can neglect the mutual friction term linear in $w_\varphi^\rns$  
on the left hand side of eqn. (\ref{sfdiff_deco}) to give
\be
\hat{\varpi} \cD_{\rm sf} \wss_\varphi
- \vec \nabla \left[ \frac{1}{{\cal R}_{\rm sf} \tilde\lambda} 
\vec\nabla \cdot(n_\rp \wns_\rP) \right] =  \frac{1}{\rho_\rp} \vec{F}_{\rm mag}^\rP
\label{sfdiff_deco2}
\ee

As in the normal case, care must be taken in extracting diffusion timescales.  For a flow with a significant non-solenoidal component, 
the pressure gradient term will dominate, leading to
\be
-\vec \nabla  \left[ \frac{1}{{\cal R}_{\rm sf} \tilde\lambda} \vec\nabla \cdot(n_\rp \wns_\rP) \right]
\approx  \frac{1}{\rho_\rp} \vec F_{\rm mag}^\rP
\label{sfdiff_ns}
\ee
We can now  estimate the diffusion timescale for the non-solenoidal flow, but now it is important to distinguish between the toroidal 
and poloidal parts: as a result of (\ref{wt1}), we should expect that the shortest diffusion timescale is associated with the toroidal 
component $\wnss_\varphi$. Indeed, specialising briefly to the weak drag case, the ratio between the toroidal and poloidal ambipolar timescales is
\be
\frac{\taunsP}{\taunsphi} =  \frac{\wnss_\varphi}{\wnss_\rP} \sim \frac{\xp}{\cB}  \gg 1 
\label{tau_ratio}
\ee
Then, combining eqns.~(\ref{wt1}) and (\ref{sfdiff_ns})  leads to (again assuming weak mutual friction) 
\be
\wnss_\varphi \sim \cR_{\rm sf} \frac{\lambda m^2 B H_{\rm c1} L}{4\pi\cB \rho^2 \xp} 
\ee
The resulting ambipolar timescale is
\be
\taunsphi \sim \frac{L}{\wnss_\varphi} \approx 6 \times 10^{23} \, \left ( \frac{10^{-17}}{\cR_{\rm sf}} \right ) 
\left (\frac{\cB}{10^{-4}}\right ) \frac{\rho_{14}^{4/3} x_{01}}{ H_{15} B_{15} T_8^6}\,~\mbox{yr} 
\label{sftau_ns}
\ee
where $H_{15} = H_{\rm c1}/(10^{15}\,\mbox{G})$, and ${\cal R}_{\rm sf} \sim 10^{-17}$ is a typical value for the superfluid suppression factor 
of eqn. (\ref{Rfits}) for $T$ close to $T_f$.  The corresponding strong mutual friction result differs only by a factor $\xp$, i.e. 
$\taunsphi \to \xp \taunsphi$.   From eqn. (\ref{wt1}), the corresponding timescale connected with poloidal flow is even longer, 
so we will not write it down.

Again in analogy with the normal case, we can consider the case of a solenoidal flow; for such a flow, eqn. 
(\ref{sfdiff_deco2}) becomes
\be
\cD_{\rm sf} \wss_\varphi  =   \frac{1}{\rho_\rp} \cF_{\rm mag}^\varpi
\label{sfdiff_sol}
\ee
where, as before, $\vec \cF_{\rm mag} / \rho_\rp$ denotes that part of the Lorentz force $\vec F_{\rm mag} / \rho_\rp$ that represents departure 
from a perfect gradient.  Again assuming only that $|\vec \cF_{\rm mag}| \lesssim |\vec{F}_{\rm mag}|$  we can make a numerical estimate of the 
corresponding ambipolar timescale:
\be
\tausphi \gtrsim  \frac{8\pi\Om \rho_\rp L^2}{B H_{\rm c1}} \approx
2 \times 10^{-3} \left ( \frac{1\,\mbox{s}}{P} \right )
\frac{ x_{01} \rho_{14} L_6^2}{H_{15} B_{15}}\, ~\mbox{s}
\label{sftau_sT} 
\ee
The timescale connected with the poloidal ambipolar diffusion can be estimated as:
\be
\tausP \sim \frac{\xp}{\cB} \tausphi \gtrsim 
2 \left ( \frac{10^{-4}}{\cB} \right )  \left ( \frac{1\,\mbox{s}}{P} \right )
\frac{L_6^2 x_{01}^2 \rho_{14}}{H_{15} B_{15}}\, ~\mbox{s}
\label{sftau_sP}
\ee
where we assumed weak mutual friction. As we pointed out earlier, the strong drag result is simply a factor $\xp$
shorter. In either case, the toroidal timescale is independent of $\cB$.  

The ambipolar diffusion timescales (\ref{sftau_ns}), (\ref{sftau_sT}), (\ref{sftau_sP}) are among the main results of this paper.
Compared to the corresponding results of the normal matter model (section~\ref{sec:normal}), these timescales are significantly
different. This is not surprising given the different nature of fluid coupling (mutual friction) in the superfluid 
model.  Indeed, these results can be understood in simple terms as follows.  
The non-solenoidal timescale is much longer in the superfluid case,  as non-solenoidal flows generate pressure and therefore compositional 
changes, which take place on a timescale set by the slow particle reactions. The solenoidal timescales are much shorter in the superfluid 
case, as the particle coupling forces which inhibit such motion are suppressed. This last point means that any solenoidal part of the 
Lorentz foce will be rapidly eliminated, leaving the star in a very slowly evolving 
transfusion-dominated equilibrium.   This contrasts with the normal fluid case, where the (lower bound) on the solenoidal timescale 
(equation (\ref{taucold2})) was of order $10^5$ years, and was therefore potentially relevant to the currently observed magnetar population.  
The implications of the obtained timescales for magnetar evolution are discussed further in section~\ref{sec:dac}.

%%%%%%%%%%%%%%%%%%%%%%%%%%%%%%%%%%%%%%%%%%%%%%%%%%%%%%%%%%%%%%

\subsection{The `hot superfluidity' regime}
\label{sec:sfhot}

We now consider the `hot' regime $ T > T_f$  where matter is superfluid but particle collisions still provide the
main coupling mechanism through the force
\be 
\vec{F}_{\rm cpl} \approx \rho_\rp \cR_\rnp D_\rnp \vec{w} 
\ee
Thus, we would expect the analysis of section~\ref{sec:normal} to be directly applicable here, 
after imposing the suppression (\ref{suppress}) and replacing the Lorentz force with the superconducting force
(\ref{Fsuper}). Recall, however, that in the case of normal matter, we were able to safely ignore the 
effects of rotation, as in the difference Euler eqn. (\ref{diff_norm}) the Coriolis force was found to be 
much smaller than the drag force. 
However, in the present superfluid case the ratio of the Coriolis to drag force is increased by a factor $\cal R_{\rm pn}$:
\be
\frac{2\Omega}{\cal D_{\rm pn} \cal R_{\rm pn}} \sim 
2 \times 10^{-16} \frac{\rho_{14}^{1/3}}{\cR_{\rm pn} T_8^2} \left(\frac{1 \, s}{P}\right)
\ee
Given that the factor $\cal R_{\rm pn}$ may be very small, it is therefore not immediately clear if rotation will be important in this 
`hot superfluidity' regime. Inspection of (\ref{Rfits}) reveals that for most of the $ T > T_f$ regime the above ratio remains small.

Then, we can readily modify the results of section \ref{sec:normal}. To begin with, the ratio $\alpha $ and the temperature $T_\alpha$ 
are modified as
\be
\alpha \to \cR_\rpn \cR_\rsf\, \alpha, \qquad T_\alpha \to ( \cR_\rpn \cR_\rsf )^{-1/8}\, T_\alpha
\label{hotmods} 
\ee
From these expressions we can safely conclude that the `hot' superfluid system will be always in the regime $T \ll T_\alpha$
where both friction-dominated and transfusion-dominated quasi-equilibria could be accessible.

The normal matter ambipolar timescales (\ref{taucold1}) and (\ref{taucold2}) are now replaced by
\bear
\nonumber \\
&& \tauns \sim \frac{4\pi\rho_\rp^2}{\cR_{\rm sf}m^2 \lambda H_{\rm c1} B} 
\nonumber \\
&& \qquad \approx 6 \times 10^{26}\, \left (\frac{10^{-17}}{\cR_{\rm sf}} \right ) 
\frac{x^2_{01} \rho_{14}^{4/3}}{T^6_8 H_{15} B_{15}}\,~ \mbox{yr} 
\label{sftau_hotns}
\\
&& \taus  \gtrsim  \frac{4\pi\rho_\rp L^2}{B H_{\rm c1}} \cR_\rnp D_\rnp 
\nonumber \\
&&\qquad \approx 9 \times 10^{-11}\, \left ( \frac{\cR_\rnp}{10^{-23}} \right) x_{01} \rho_{14}^{2/3} 
\frac{  L_6^2 T^2_8}{H_{15} B_{15}}\,~ \mbox{s}
\label{sftau_hots}
\eear
where ${\cal R}_{\rm np} \sim 10^{-23}$ is a typical value for the collisional suppression factor for temperatures close to $T_f$, 
as estimated using eqn. (\ref{Rfits}). As it was the case in the `cold' superfluid regime, the evolution of the solenoidal and 
non-solenoidal flows in the `hot' regime are associated, respectively, with a very short and very long timescale.
Thus, the system is expected to be in a transfusion-dominated quasi-equilibrium.

Finally we check if the condition $ w \gg w_\rpe$ is satisfied. It is straightforward to show that this is the case for both
regimes of superfluidity. For type II proton superconductivity the magnetic field appearing in the Amp\`ere law  
is the so-called London field $\vec b_{\rm L}$ \citep{clp00,supercon}.  The London field itself is given by the standard expression 
(see \citet{tilley} for details),
\be
\vec{b}_{\rm L} = -\frac{2mc}{e} \vec{\Om} \quad \Rightarrow \quad b_{\rm L} \approx 10^{-3} \left ( \frac{1\,\mbox{s}}{P} \right ) \,\mbox{G}
\ee
The intrinsic weakness of the London field implies a relative proton-electron velocity $w_\rpe$ many orders of 
magnitude smaller than the numerical estimate (\ref{wpe}). Related to this last point, the enormous weakening of the 
electric current $ \vec j = e n_\rp \vec w_\rpe$ in the presence of type II proton superconductivity also implies
a negligible magnetic field evolution due to the Hall drift.

%%%%%%%%%%%%%%%%%%%%%%%%%%%%%%%%%%%%%%%%%%%%%%%%%%%%%%%%%%%%%%%%%

\section{Discussion: implications for magnetars}
\label{sec:dac}

We now discuss the main results of this paper, consisting of the ambipolar diffusion timescales  (\ref{sftau_ns}), (\ref{sftau_sT}), 
(\ref{sftau_sP}) for the `cold' superfluid regime and (\ref{sftau_hotns}), (\ref{sftau_hots}) for the `hot' superfluid regime, in the context 
of magnetic field evolution in magnetars. Ambipolar diffusion could provide a viable mechanism for magnetic field evolution in the observed 
magnetar population provided the associated timescale is comparable to the estimated ages  $\sim 10^3-10^4\,\mbox{yr} $ for these objects. 

Let us first examine the non-solenoidal timescales of eqns. (\ref{sftau_ns}) and (\ref{sftau_hotns}). As a consequence of the superfluid 
suppression of particle reactions (represented by the exponentially small factor $\cR_\rsf$) all non-solenoidal timescales become extremely long, 
much longer than the corresponding normal matter timescales (\ref{tauhot}), (\ref{taucold1}) and the estimated magnetar ages. An interesting 
exception could be the `cold' superfluid timescale (\ref{sftau_ns}) if the mutual friction coupling could be reduced at the level of 
$\cB \sim 10^{-4}\, \cR_\rsf$. However, such a very small value of ${\cal B}$ is unlikely to occur if mutual friction is weak, given that 
the rather well understood electron scattering by neutron vortices leads to a much stronger coupling 
$\cB \sim 10^{-5} - 10^{-4}$ \citep{als88,na06}. In principle, one could have a very small $\cB$ in the opposite limit of nearly perfect 
vortex pinning. However, in the absence of input from microphysics, $\cB$ would have to be arbitrarily chosen to be of the order $\sim 10^{-19}$ 
in order for our model to produce ambipolar timescales of the order of $10^4\,\mbox{yr}$. Regardless of this issue of `fine-tuning', and as we 
pointed out earlier, the regime of vortex pinning would require a more sophisticated superfluid model with detailed vortex/fluxtube dynamics 
(e.g. the superconducting model of \citet{supercon}).
 
A similar situation emerges when we examine the solenoidal timescales (\ref{sftau_sT}), (\ref{sftau_sP}) and (\ref{sftau_hots}). 
They clearly show that any solenoidal component of the Lorentz force will be eliminated on a very short dynamical-like timescale, 
much shorter than the estimated ages of the magnetar population.  This means that the superfluid case is very much simpler than the normal 
fluid case, where the (lower bound) on the solenoidal timescale (equation (\ref{taucold2})) was comparable to the estimated magnetar ages, 
so that, in the normal fluid regime,  it was not clear if such flows were physically relevant. 

Thus, we conclude that ambipolar diffusion is unlikely to be the driving mechanism for magnetic field evolution in the observed magnetars 
{\it unless} we invoke mutual friction very much weaker than current modelling suggests.   This does not sit well with the standard magnetar 
model, where magnetic field diffusion is needed to account for the observed activity of the SGRs and AXPs.  There do, however, exist mechanisms 
alternative to core ambipolar diffusion that might supply the required field evolution.  One possibility is  a combination of Hall and Ohmic 
processes in the stellar crust \citep{pmg09}.  More speculatively, it is possible that vortex-fluxtube interactions, not included in our model, 
might provide the crucial evolution mechanism; such a process will be considered elsewhere. Yet another possibility is that the inner core is 
composed of exotic matter such as hyperons or quark condensates to which our model does not apply. Thus the poorly constrained physics of the inner 
core may allow for processes that lead to magnetic evolution on the required timescale. 

It is possible to consider situations where only one of the proton and neutron fluids form condensates without resorting to further calculation.  
Let us first consider the case where the neutrons are superfluid but 
the protons normal.  This would apply e.g. if the magnetic field were to exceed the `second critical field' above which proton superconductivity 
is suppressed, i.e. if $B > H_{{\rm c}2} \sim 10^{16}$ G \citep{tilley}.  In such a regime the absence of entrained protons around the neutron vortices would suppress the mutual friction coupling force \citep{feibe71, sss82}, so that the hot superfluid regime considered in section \ref{sec:sfhot} would be relevant.  The lack of proton superfluidity would also have the effect of decreasing the importance of the superfluid suppression on the particle collision and reaction rates, so the suppression factors in this case would lie somewhere between the values given in eqn. (\ref{Rfits}) and unity \citep{haensel00,haensel01,baiko01}.  From eqns. (\ref{sftau_hotns}) and (\ref{sftau_hots}) we see this would have the effect of pushing both the solenoidal and non-solenoidal diffusions timescales in the direction of physical interest.

Now consider the case where the neutrons are normal but the protons superconducting.  Such a model formed the basis of the analysis of \citet{act04}, 
who studied the link between magnetic field decay and thermal emission.  In terms of our analysis, the lack of neutron superfluidity eliminates the mutual 
friction force, so that we can simply apply the `hot superfluidity' model of section \ref{sec:sfhot}, where the mutual friction 
force was not absent but was negligibly small. We can therefore make direct use of eqns. (\ref{sftau_hotns}) and (\ref{sftau_hots}), with suitable values 
of the suppression factors \citep{baiko01,haensel01}. We find that the qualitative picture is unchanged, with the non-solenoidal timescale remaining much 
too long to be of physical interest, while the (lower bound) on the solenoidal timescale remains extremely short.

Our model has a number of obvious weaknesses, the most serious of which is the neglect of interactions between the neutron vortices and the 
magnetic fluxtubes. The inclusion of such effects would provide a logical extension of the analysis of this paper.
Our analysis could also be improved by considering particular realistic field configurations.  This would allow explicit calculations of
how the magnetic field evolves, giving more accurate timescales, and information on how the geometry of the field evolves in time.  
Indeed, such calculations have been carried out recently by \citet{hoyos08,hoyos10} for  non-superfluid stars.

Related to this last point, our work illustrates the need for the construction of accurate axisymmetric equilibria, to act as backgrounds about 
which to monitor the field evolution.  However, as we have endeavored to point out, the construction of such solutions in complicated by the 
simultaneous existence of magnetic forces 
and particle reactions: the magnetic stresses imply the existence of a non-zero chemical imbalance, which in turn implies the existence of 
particle reactions, relative motion of the species, and particle scattering. 
Interestingly, in the case of normal matter, our analysis in section \ref{sec:normal} makes clear that, if the $\vec \cF_{\rm mag}/\rho_\rp$ 
component of the Lorentz force (the part which is not the gradient of a scalar) is significant, then the friction-dominated equilibrium that 
results has particle scattering as an important component  of the force balance.  Such forces have been absent from all magnetic equilibrium 
calculations to date, both perturbative (e.g. \citet{haskell08,ciolfi09,sam09}) and numerical (e.g. \citet{bn06}), and so these normal fluid 
calculations might be significantly in error.  Whether such a component of the Lorentz force is relevant, or whether it is removed on some fast 
dynamical timescale by processes not considered here, is certainly an issue affecting the normal fluid case.  Clearly, a scheme 
for producing pseudo-stationary equilibria needs to be developed, for both normal matter and superfluid/superconducting stars, which would 
then provide a background for more detailed investigations of secular field evolution.

%%%%%%%%%%%%%%%%%%%%%%%%%%%%%%%%%%%%%%%%%%%%%%%%%%%%

\section*{Acknowledgments}
The authors are grateful to Nils Andersson, Jos\'e Pons, Chris Thompson and the referee, Andreas Reisenegger, 
for valuable comments on our manuscript.
KG is supported by an Alexander von Humboldt fellowship and by the German Science Foundation (DFG) via SFB/TR7.
DIJ is supported by STFC via grant number PP/E001025/1. LS is supported by the European Research Council under Contract 
No. 204059-QPQV, and the Swedish Research Council under Contract No. 2007-4422. 
The authors also acknowledge support from COMPSTAR (an ESF Research Networking Programme).

%%%%%%%%%%%%%%%%%%%%%%%%%%%%%%%%%%%%%%%%%%%%%%%%%%%

\end{document}